\documentclass{mem}
\usepackage{natbib}
\usepackage{txfonts}
\usepackage{balance}
\usepackage{graphicx}
\usepackage[a4paper]{hyperref}
\idline{79}{1}
\begin{document}

\title{
Physics of the central region in the quasar 0850+581
}

\subtitle{}

\author{
Y.~Y.~Kovalev\inst{1,2},
A.~P.~Lobanov\inst{1},
\and
A.~B.~Pushkarev\inst{1,3,4}
}

\offprints{Y.~Y.~Kovalev}

\institute{
Max-Planck-Institut f\"ur Radioastronomie,
Auf dem H\"ugel 69, 53123 Bonn, Germany\\
\email{ykovalev,alobanov,apushkar@mpifr-bonn.mpg.de}
\and
Astro Space Center of Lebedev Physical Institute, Profsoyuznaya 84/32, Moscow 117997, Russia 
\and
Crimean Astrophysical Observatory, Nauchny, Crimea, Ukraine
\and
Pulkovo Observatory, Pulkovskoe highway 65/1, St. Petersburg 196140, Russia
}

\authorrunning{Kovalev et al.}
\titlerunning{Central parsecs in the quasar 0850+581}

\abstract{
The apparent position of the origin (core) of extragalactic radio jets
shifts with the observing frequency, owing to synchrotron
self-absorption and external absorption. One of the largest core shifts was
detected by us in the quasar 0850+581 between 2 and 8~GHz. We have
followed this up recently by a dedicated VLBA experiment at 5, 8, 15,
24, and 43~GHz. First results from this study enabled estimating the
absolute geometry and physical conditions in the parsec-scale jet origin.

}
\maketitle{}

\section{Introduction}

In VLBI images of relativistic jets, the location of the narrow end of
the jet (branded the ``core'') is fundamentally determined by absorption
in the radio emitting plasma itself (synchrotron self-absorption) and/or
in the material surrounding the flow
\citep{BlandfordKonigl79,Koenigl81,L98} and can be further modified by
strong pressure and density gradients in the flow \citep{L98}. At any
given observing frequency, $\nu$, the core is located in the jet region
with the optical depth $\tau_s(\nu)\approx 1$, which causes its absolute
position, $r_\mathrm{c}$, to shift $\propto \nu^{-1/k_\mathrm{r}}$. If
the core is self-absorbed and in equipartition, $k_\mathrm{r}=1$
\citep{BlandfordKonigl79}; $k_\mathrm{r}$ can be larger in the presence
of external absorption or pressure/density gradients in the flow
\citep{L98}.

Changes of the core position measured between three or more frequencies
can be used for determining the value of $k_\mathrm{r}$, estimating the
strength of the magnetic field in the nuclear region and the offset of
the observed core positions from the true base of the jet \citep{L98}.
The power index $k_\mathrm{r}$ itself can vary with frequency due to
pressure and density gradients or absorption in the surrounding medium,
most likely, associated with the broad-line region.

If the core shifts and $k_\mathrm{r}$ are measured between four, or
more, frequencies, the following can be addressed in detail.
The magnetic field distribution can be reconstructed in the
ultra-compact region of the jet and estimates of the total (kinetic plus
magnetic field) power, the synchrotron luminosity, and the maximum
brightness temperature, $T_\mathrm{b,max}$ in the jet can be made.
In addition, the ratio of particle energy and magnetic field energy can
be estimated from the derived $T_\mathrm{b,max}$. This would
enable testing the \cite{Koenigl81} model and several of its later
modifications \citep[e.g.,][]{HM86,BM96}.
The location of the central engine and the geometry of the jet can be
determined. Estimation of the distance from the nucleus to the jet
origin will enable constraining the self-similar jet model
\citep{Marscher95} and the particle-cascade model \citep{BL95}.

In this paper we present first results of a dedicated high resolution
VLBA study of the core shift effect in a central region of the distant
luminous quasar 0850+581.

\section{A large core shift in the quasar 0850+581}

We have imaged the NRAO archival data and used results of the VLBA S/X
project BF\,025 to estimate core shifts in a number of sources \citep[see
description of the BF\,025 program in][]{FC00}. We have found an
intriguing case of the distant quasar 0850+581 (redshift $z=1.318$, SDSS
release~2) for which the core shift was estimated to be 1.5~mas between
2.3 and 8.6~GHz \citep{Kovalev_cs_2008}. The total error of the core
shift value for this object is dominated by blending of the first bright
jet component with the core at long wavelengths. However, the distance
between the core and this component was about 0.5~mas. This means that
the huge core shift value about 1~mas measured between 2.3 and 8.6~GHz
and 2.3 and 15~GHz must be real. 

So far, this is one of the largest core shifts detected. This object
is particularly suited for VLBI observations at high radio frequencies
because of the bright component located close to the core (see
Figure~\ref{f:0850images}). We observed this source in a dedicated
5--43~GHz VLBA experiment to confirm the core shift and study
physics of the nuclear region in this quasar.

\section{5-43 GHz VLBA observations and core shift measurements results}

\begin{figure*}[p]
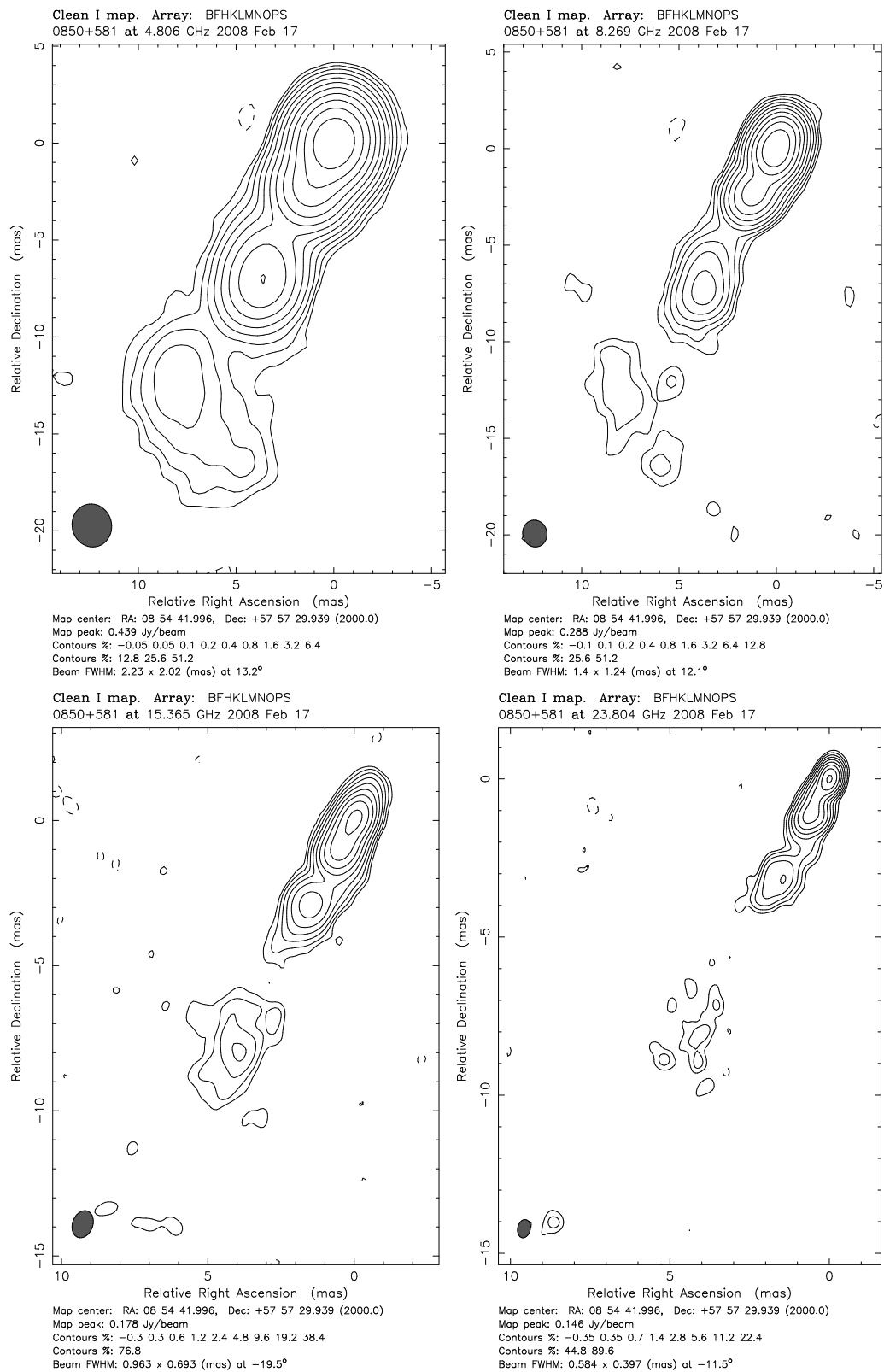

\resizebox{\hsize}{!}{
   \includegraphics[clip=true,trim=0cm 0cm 0cm -0.5cm]{kovalev_f1a.eps}
   \includegraphics[clip=true,trim=-0.5cm 0cm 0cm -0.5cm]{kovalev_f1b.eps}
}
\resizebox{\hsize}{!}{
   \includegraphics[clip=true,trim=0cm 0cm 0cm -0.5cm]{kovalev_f1c.eps}
   \includegraphics[clip=true,trim=-0.5cm 0cm 0cm -0.5cm]{kovalev_f1d.eps}
}
\caption{
\label{f:0850images}
\footnotesize
Stokes I contour CLEAN maps of the quasar 0850+581
observed quasi-simultaneously by VLBA on February~17, 2008.
Images are restored with natural weighting.
One milliarcsecond is about 8.4~pc.
}
\end{figure*}

We performed VLBA observations of the quasar 0850+581
at 5, 8, 15, 24, \& 43 GHz frequency bands during 15~hours on February
17, 2008 (project code BK\,142).
This was done as a phase referencing experiment, which includes a 
phase calibrator selected from the VLBA Calibrator Survey
\citep[e.g.,][]{VCS6} allowing to do differential
astrometry measurements.
Imaging results at four bands with comparable dynamic range
are presented in Figure~\ref{f:0850images}.

First results of the core-shift measurements are presented in
Figure~\ref{f:0850cs}. They include one data point at 2.3~GHz, which is
an extrapolation from previous measurements \citep{Kovalev_cs_2008}. It
should be noted that these results are achieved using the
self-referencing alignment method and spectral~index analysis (see for
details \citealt{Kovalev_cs_2008}). We plan to perform a differential
astrometry analysis at a later point and present results elsewhere.

\section{Physical properties of the jet origin}

\begin{figure}[t]
\resizebox{\hsize}{!}{\includegraphics[angle=-90,clip=true,trim=0cm 0cm 0cm 0cm]{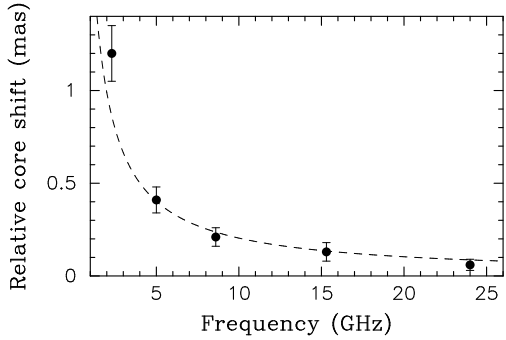}}
\caption{
\label{f:0850cs}
\footnotesize
Frequency dependence of the core shift value measured relative to the
core position at 43 GHz. The curve represents the best fit for the
function $r_\mathrm{c}\propto\nu^{-k_\mathrm{r}}$, where
$k_r=1.1\pm0.1$.
}
\end{figure}

We did not detect significant deviation of $k_\mathrm{r}$ from unity 
(Figure~\ref{f:0850cs}), which mean that the shift happens purely as a
result of synchrotron self-absorption. External absorption is not
significant at the observed core positions at least until 24~GHz.

We apply the method suggested by \cite{L98} and assume the following
parameters for the jet of 0850+581: Lorentz factor $\Gamma=10$ (from
kinematics measurements by \citealt{2cmPaperIII}), viewing angle
$10^\circ$ (from the two-sided kiloparsec scale morphology by
\citealt{Reid_etal95}), jet opening angle $1/\Gamma^2=0.6^\circ$. The
following parameters were estimated. Distance from the observed core to
the central supermassive black hole is found to differ from 17~pc (at
5~GHz) to 5~pc (at 24~GHz). Magnetic field strength at a distance of
1~pc from the nucleus is $B=3.1\pm0.2$~G (for assumed particle density
1000~cm$^{-3}$), which is consistent with equipartition magnetic field
($B=2.7\pm0.7$~G). Magnetic field strength at the position of the
apparent 24~GHz core is $B=0.2\pm0.4$~G.

\section{Summary}

The nuclear opacity in relativistic jets significantly affects observed
positions of compact radio cores. This effect provides an efficient tool
to study physics of compact jet nuclei. Application of this to the
quasar 0850+581 with a large core shift effect allowed to determine
geometry and physics in the inner region of compact jet.

The distant quasar 0850+581 is an ICRF source \citep{ICRF98} and is
monitored in astrometric/geodetic VLBI sessions. Results of this study
can be used to improving the absolute position of the object and for
aligning positions measured at different observing bands.

\begin{acknowledgements}
The National Radio Astronomy Observatory is a facility of the National
Science Foundation operated under cooperative agreement by Associated
Universities, Inc. We thank D.~Vir~Lal for careful reading the
manuscript and useful comments. Y.~Y.~Kovalev is a Research Fellow of
the Alexander von Humboldt Foundation. Y.~Y.~Kovalev was supported in
part by the Russian Foundation for Basic Research (project 05-02-17377,
08-02-00545). This research has made use of NASA's Astrophysics Data
System and the NASA/IPAC Extragalactic Database (NED) which is operated
by the Jet Propulsion Laboratory, California Institute of Technology,
under contract with the National Aeronautics and Space Administration.
\end{acknowledgements}

\bibliographystyle{aa}
\bibliography{yyk}

\begin{thebibliography}{13}
\expandafter\ifx\csname natexlab\endcsname\relax\def\natexlab#1{#1}\fi

\bibitem[{{Blandford} \& {K\"onigl}(1979)}]{BlandfordKonigl79}
{Blandford}, R.~D. \& {K\"onigl}, A. 1979, \apj, 232, 34

\bibitem[{{Blandford} \& {Levinson}(1995)}]{BL95}
{Blandford}, R.~D. \& {Levinson}, A. 1995, \apj, 441, 79

\bibitem[{{Bloom} \& {Marscher}(1996)}]{BM96}
{Bloom}, S.~D. \& {Marscher}, A.~P. 1996, \apj, 461, 657

\bibitem[{{Fey} \& {Charlot}(2000)}]{FC00}
{Fey}, A.~L. \& {Charlot}, P. 2000, \apjs, 128, 17

\bibitem[{{Hutter} \& {Mufson}(1986)}]{HM86}
{Hutter}, D.~J. \& {Mufson}, S.~L. 1986, \apj, 301, 50

\bibitem[{{Kellermann} {et~al.}(2004){Kellermann}, {Lister}, {Homan},
  {Vermeulen}, {Cohen}, {Ros}, {Kadler}, {Zensus}, \& {Kovalev}}]{2cmPaperIII}
{Kellermann}, K.~I., {Lister}, M.~L., {Homan}, D.~C., {et~al.} 2004, \apj, 609,
  539

\bibitem[{{K\"onigl}(1981)}]{Koenigl81}
{K\"onigl}, A. 1981, \apj, 243, 700

\bibitem[{{Kovalev} {et~al.}(2008){Kovalev}, {Lobanov}, {Pushkarev}, \&
  {Zensus}}]{Kovalev_cs_2008}
{Kovalev}, Y.~Y., {Lobanov}, A.~P., {Pushkarev}, A.~B., \& {Zensus}, J.~A.
  2008, \aap, 483, 759

\bibitem[{{Lobanov}(1998)}]{L98}
{Lobanov}, A.~P. 1998, \aap, 330, 79

\bibitem[{{Ma} {et~al.}(1998){Ma}, {Arias}, {Eubanks}, {Fey}, {Gontier},
  {Jacobs}, {Sovers}, {Archinal}, \& {Charlot}}]{ICRF98}
{Ma}, C., {Arias}, E.~F., {Eubanks}, T.~M., {et~al.} 1998, \aj, 116, 516

\bibitem[{{Marscher}(1995)}]{Marscher95}
{Marscher}, A.~P. 1995, Proceedings of the National Academy of Science, 92,
  11439

\bibitem[{{Petrov} {et~al.}(2008){Petrov}, {Kovalev}, {Fomalont}, \&
  {Gordon}}]{VCS6}
{Petrov}, L., {Kovalev}, Y.~Y., {Fomalont}, E.~B., \& {Gordon}, D. 2008, \aj,
  136, 580

\bibitem[{{Reid} {et~al.}(1995){Reid}, {Shone}, {Akujor}, {Browne}, {Murphy},
  {Pedelty}, {Rudnick}, \& {Walsh}}]{Reid_etal95}
{Reid}, A., {Shone}, D.~L., {Akujor}, C.~E., {et~al.} 1995, \aaps, 110, 213

\end{thebibliography}

\end{document}